\documentclass{moriond}
\usepackage{amssymb}
\usepackage{amsmath}
\usepackage{caption}

\def\be{\begin{equation}}
\def\ee{\end{equation}}
\def\bea{\begin{eqnarray}}
\def\eea{\end{eqnarray}}

\newcommand{\Photo}{}
\newcommand{\lemaitre}{\textsc{Lema\^itre}}

\begin{document}
\vspace*{4cm}
\title{Development of an ultra-fast, likelihood-based, distance inference framework for the next generation of Type Ia Supernova Surveys}

\author{ Dylan KUHN, Marc BETOULE, on behalf of the \lemaitre \ collaboration }

\address{Sorbonne Université, CRNS, Université de Paris, LPNHE, 75252 Paris Cedex 05, France}

\maketitle\abstracts{In this work, we present \textsc{EDRIS} (French for \emph{Distance Estimator for Incomplete Supernova Surveys}), a cosmological inference framework tailored to reconstruct unbiased cosmological distances from type Ia supernovae light-curve parameters. This goal is achieved by including data truncation directly in the statistical model which takes care of the standardization of luminosity distances. It allows us to build a single-step distance estimate by maximizing the corresponding likelihood, free from the biases the survey detection limits would introduce otherwise. Moreover, we expect the current worldwide statistics to be multiplied by O(10) in the upcoming years. This provides a new challenge to handle as the cosmological analysis must stay computationally towable. We show that the optimization methods used in \textsc{EDRIS} allow for a reasonable time complexity of O($N^2$) resulting in a very fast inference process (O(10s) for 1500 supernovae).}

\section{Introduction}

The main goal of \textsc{EDRIS} is to handle the instrumental bias known as Malmquist bias. The observable magnitude limitation inherent to each survey induces a preferential observation of the intrinsically brighter objects and, therefore, a negative bias on the distance estimator\cite{teerikorpi98}. The usual way to prevent this estimator from being biased is to run extensive simulations to compute a bias correction which is propagated to the reconstructed distances afterwards\cite{betoule14,kessler17}. However, this method does not scale well with the number of data and we expect the current worldwide statistics to increase tenfold in the upcoming years. To tackle both the bias issue and the scaling issue, we propose an innovative approach based on a truncated likelihood minimization. This allows to integrate a modeling of the selection effect in the distance standardization model. The presented method is implemented in the context of a large data combination named \lemaitre. The \lemaitre \ analysis is an end-to-end cosmology analysis using three unpublished SNe Ia samples (ZTF, SNLS 5y, HSC/Subaru), totalizing O(4000) supernovae. All data are processed with a common pipeline going from pixels to cosmological inference, this last role being filled by \textsc{EDRIS}.

\section{Modeling of the Malmquist bias}

The common model used to describe the behaviour of type Ia supernovae is the Tripp model\cite{tripp98} (see equation \ref{eq:tripp}). The usual standardization process involves two parameters (color and stretch) but this model can easily be generalized to consider an arbitrary number of standardization parameters. The $^*$ symbols indicate latent parameters we need to introduce to properly account for the fact that the $c$ and $x_1$ measurements are affected with uncertainties that cannot be neglected.

\begin{equation}\label{eq:tripp}
    Y_i = \begin{pmatrix}
        m_i \\
        x_{1,i} \\
        c_i
    \end{pmatrix} = \begin{pmatrix}
        M + \mu_i(z, \theta) + \alpha x_{1,i}^* + \beta c_i^* \\
        x_{1,i}^* \\
        c_i^*
    \end{pmatrix} + \begin{pmatrix}
        \epsilon_i \\
        0 \\
        0
    \end{pmatrix} \ \mathrm{with} \ \epsilon_i \sim \mathcal{N}(0, \sigma_{int}^2)
\end{equation}

$m_i$, $x_{1,i}$, $c_i$ and $\mu_i$ respectively refer to the apparent magnitude of the supernova $i$, its stretch parameter, its color parameter and its distance modulus. $M$ is the absolute magnitude of the supernovae and $\sigma_{int}$ the absolute magnitude dispersion of the SNe Ia population. The $(m_i, x_{1,i}, c_i)$ parameters are correlated and we denote the full covariance matrix $C$. The truncation effect can then be written as:

\bea
    & &Y_i^{obs} = Y_i + \eta_i \ \mathrm{if} \ m_i \leqslant m_{lim} + \kappa_i \ \mathrm{with} \ \eta_i \sim \mathcal{N}(0, C_i) \ \mathrm{and} \ \kappa_i \sim \mathcal{N}(0, \sigma_d^2) \nonumber \\
    & &Y_i^{obs} \ \mathrm{is \ unobserved \ otherwise}
\eea

where $\eta_i$ is the measurement noise, $m_{lim}$ is the absolute magnitude of the survey and $\kappa_i$ is its fluctuation due to the variability in observation conditions $\sigma_d$. The negative log-likelihood function associated to the model described above is based on the standard likelihood associated to the multivariate normal distribution. The novelty lies in the sum in equation \ref{eq:edrislikelihood}. Starting from the Bayes theorem, we derived two new terms for the likelihood. Both depend on the cumulative distribution function of the normal distribution and the selection function parameters, allowing to take into account the truncation of data\cite{rubin15}.

\bea\label{eq:edrislikelihood}
	\Gamma &=& -\ln(|W|) + r^{\dagger}Wr \nonumber \\
           &+& \sum_i 2 \ln \left( \Phi \left( \frac{m_{lim}- M - \mu_i - \alpha x^*_{1,i} - \beta c^*_i}{\sqrt{\sigma_{int}^2 + \sigma_d^2}} \right) \right) -2\ln \left( \Phi \left( \frac{m_{lim} - m_i^{obs}}{\sqrt{\sigma_d^2 + f(C_i)}} \right) \right)
\eea

with $\Phi(z) = \frac{1}{2} \left(1 + \mathrm{erf} \left(\frac{z}{\sqrt{2}} \right) \right)$, $W = C^{-1}$ and $r = \begin{pmatrix}
    m^{obs} \\
    x_1^{obs} \\
    c^{obs}
\end{pmatrix} - \begin{pmatrix}
    M^* - \mu - \alpha x_1^* - \beta c^* \\
    x_1^* \\
    c^*
\end{pmatrix}$.

The modeling of the Malmquist bias described is equivalent to the modeling of the selection functions by three independant sigmoids (one per survey). Each sigmoid is characterized by a central value ($m_{lim}$) and a width ($\sigma_d$).

\section{Estimation of distances and cosmological parameters}

\begin{figure}[htbp]
\begin{minipage}[t]{0.48\linewidth}
\includegraphics[width=\linewidth, keepaspectratio=true]{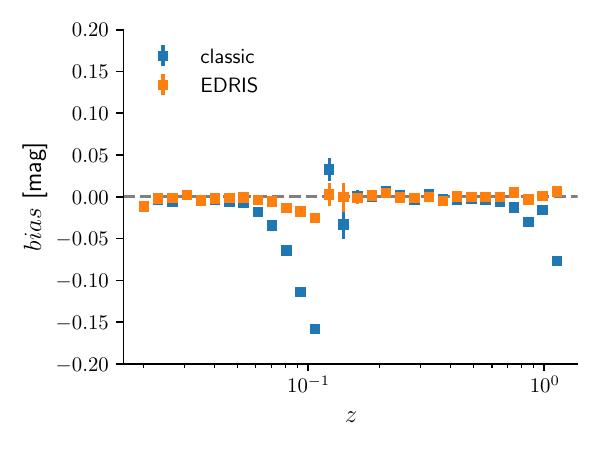}
\end{minipage}
\hfill
\begin{minipage}[t]{0.48\linewidth}
\includegraphics[width=\linewidth, keepaspectratio=true]{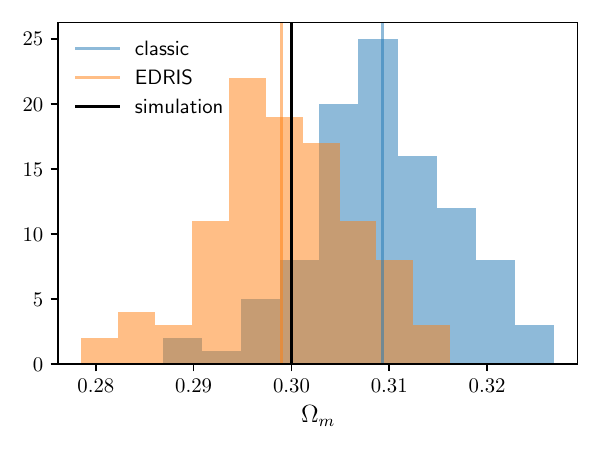}
\end{minipage}
\caption[]{\emph{Left panel:} Bias on the reconstructed binned distances moduli for each estimator. \emph{Right panel:} Histogram of the reconstructed matter energy density parameter for each estimator.}
\label{fig:biasedris}
\end{figure}

In statistics, the estimator of the variance (including $\sigma_{int}$ in our case) is biased. As a consequence, \textsc{EDRIS} is intrinsically biased. To quantify this effect, we performed 100 Monte-Carlo simulations. The simulated SNe Ia light-curves and redshifts are generated from observation logs with \textsc{SkySurvey}\footnote{\url{https://skysurvey.readthedocs.io/en/latest/}}. For this analysis, we consider a simplified case with only one standardization parameter (colour). This allows to fit the light-curves with a 1D version of \textsc{SALT}\cite{hazenberg19}. Thus, we keep the complexity of our problem (the light-curve model is still trained) while running the analysis faster. Simulations parameters are presented in table \ref{tab:simuparams} while results are presented in figure \ref{fig:biasedris}. The left panel shows the mean of the difference between the reconstructed distance and the simulation input in 30 bins logarithmically distributed in redshift. The blue points correspond to the classic maximum likelihood estimator for multivariate normal distribution and show a strong negative bias on reconstructed distances due to the selection effect for each survey. The orange points correspond to the distance estimator including the modeling of the selection effect and show no significant bias on $\Omega_m$ (see table \ref{tab:biasom}).

\begin{table}[htbp]
\caption[]{Simulations parameters for the Monte-Carlo analysis.}
\label{tab:simuparams}
\vspace{0.4cm}
\begin{center}
\begin{tabular}{||c c c c c c c||} 
 \hline
 Survey & Redshift range & $\beta$ & $M$ & Cosmology & $m_{lim}$ & $\sigma_d$ \\ [0.5ex] 
 \hline\hline
 ZTF & $[0.0, 0.2]$ & $3.15$ & $-19.0$ & Planck 2018\cite{planck18} & $18.59$ & $0.1$ \\ 
 \hline
 SNLS 5y & $[0.0, 1.07]$ & $3.15$ & $-19.0$ & Planck 2018\cite{planck18} & $24.38$ & $0.25$ \\
 \hline
 HSC/Subaru & $[0.0, 1.6]$ & $3.15$ & $-19.0$ & Planck 2018\cite{planck18} & $25.2$ & $0.07$ \\
 \hline
\end{tabular}
\end{center}
\end{table}

\begin{table}[htbp]
\caption[]{Bias and uncertainty on $\Omega_m$ derived from Monte-Carlo simulations for both estimators.}
\label{tab:biasom}
\vspace{0.4cm}
\begin{center}
\begin{tabular}{||c c c c||} 
 \hline
 Estimator & Bias on $\Omega_m$ & Mean of reconstructed $\Omega_m$ & $\sigma(\Omega_m)$ \\ [0.5ex] 
 \hline\hline
 classic & 0.0093 $\pm$ 0.0007 & 0.309 & 0.007 \\ 
 \hline
 EDRIS & -0.0010 $\pm$ 0.0007 & 0.299 & 0.007 \\
 \hline
\end{tabular}
\end{center}
\end{table}

\section{Acceleration of the computation}

As the covariance matrix of the observations depends on the estimated $\sigma_{int}$ parameter, we need to invert it at each step of the likelihood minimization. However, this dependency is simple enough to take advantage of the Schur complement technique\cite{ouellette81}. The first step consists in taking into account the block structure of $C$ (see equation \ref{eq:decompositionW}) and writing $r$ as $(r_1, r_2)$ to match this structure. 

\begin{equation}\label{eq:decompositionW}
    W = \begin{pmatrix}
    C_{mm} + \sigma_{int}^2 I_N & C_1 \\
    C_1^{\dagger} & C_2
    \end{pmatrix}^{-1}
\end{equation}

Then, we can compute and diagonalize (see equation \ref{eq:SchurDiagonalization}) the Schur complement $S = C_{mm} + \sigma_{int}^2I_N - C_1C_2^{-1}C_1^{\dagger}$ of the lower right block $C_2$ in $C$.

\begin{equation}\label{eq:SchurDiagonalization}
    S^{-1} = Q(\Lambda + \sigma_{int}^2 I_N)^{-1}Q^{\dagger}
\end{equation}

The two first terms of the likelihood can then be written:

\begin{equation}
    -\ln(|W|) = \sum_i \ln(\Lambda_i + \sigma_{int}^2) + \ln(|C_2|)
\end{equation}

\begin{equation}\label{eq:chi2}
    r^{\dagger}Wr = r_1^{\dagger}S^{-1}r_1 
                    -2 r_1^{\dagger}S^{-1}C_1C_2^{-1}r_2 
                    + r_2^{\dagger}C_2^{-1}r_2 
                    + r_2^{\dagger}C_2^{-1}C_1^{\dagger}S^{-1}C_1C_2^{-1}r_2
\end{equation}

\begin{figure}[htbp]
\begin{center}
\begin{minipage}[t]{0.48\linewidth}
\includegraphics[width=\linewidth, keepaspectratio=true]{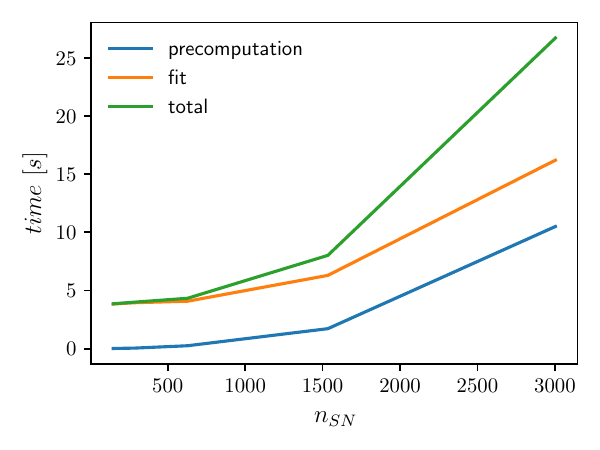}
\end{minipage}
\end{center}
\caption[]{Time scaling of the likelihood minimization as a function of the number of simulated supernovae. The blue, orange and green lines respectively account for the precomputation of constant terms, the cosmological inference and the total time.}
\label{fig:scaling}
\end{figure}

The last step consists in precomputing every constant matrix-matrix products in equation \ref{eq:chi2}. Thus, only matrix-vector products remain, which allows to scale the likelihood computation in O(N$^2$) instead of O(N$^3$) when working with naive determinant computation and matrix inversion. To push the optimization even further, we decided to implement a hessian-free minimization method using \textsc{JAX}\footnote{\url{https://jax.readthedocs.io/en/latest/}}. The total computing time is presented in figure \ref{fig:scaling}. We manage to reach a minimization in O(10s) for 1500 supernovae, including the standardization of the magnitudes, the estimation of $\sigma_{int}$ and the correction of the Malmquist bias.

\end{document}